\magnification=1200
\hsize=6truein
\vsize=8.5truein
\def\a{\alpha}

\def\r{\rho}
\def\fm{{\rm fm}}

\def\O{\Omega} 
\def\he4{{}^4{\rm He}}

\def\qq{\quad}
\def\tablerule{\noalign{\hrule}}
\def\htm{{\hbar^2 \over m}}

\def\b{\beta}
\def\P{{\rm\Psi}}
\def\R{{\widetilde R}}
\def\aa{{\alpha,\alpha'}}
\def\bb{{\beta,\beta'}}
\def\kk{{K,K'}}
\def\ap{{\alpha'}}
\def\bp{{\beta'}}
\def\Y{{\cal Y}}
\def\x{{\bf x}}
\def\y{{\bf y}}
%
\def\yx{Y_{l_\a}({\hat x}_i)}

\def\yyl{Y_{L}({\hat y}_i)}
\vglue 2.0cm
\centerline{\bf STUDY OF BOUND AND SCATTERING STATES }
\centerline{\bf OF THREE--NUCLEON SYSTEMS}
\centerline{}
\centerline{A. Kievsky, M. Viviani}\smallskip
\centerline{{\sl INFN, Sezione di Pisa, Piazza Torricelli 2, 56100 Pisa,
Italy}}\medskip
\centerline{S. Rosati}\smallskip
\centerline{{\sl Dipartimento di Fisica, Universita' di Pisa,
Piazza Torricelli 2,  56100 Pisa, Italy}}\smallskip
\centerline{{\sl INFN, Sezione di Pisa, Piazza Torricelli 2, 56100 Pisa, 
Italy}}
\bigskip

\noindent{{\bf Abstract.}} A variational technique to
describe the ground and scattering states below the break--up
threshold for a three-nucleon system is developed.
The method consists in 
expanding the wave function in terms of correlated Harmonic
Hyperspherical functions suitable to handle the large repulsion contained
in the nuclear potential at short distances;
three body forces have also been considered.
The inclusion of the pp Coulomb repulsion in the p--d processes does not
cause any particular problem,
since no partial wave decomposition of the interaction is performed.
Accurate numerical results are given for
ground state properties and scattering lenghts, phase shifts and mixing
parameters at three different energies of the incident nucleon. The
agreement with other available results and with experimental analyses is
higlly satisfactory.
\bigskip
\noindent{{\bf 1. Introduction}}
\medskip

A sophisticated variational technique to describe the bound state of a 
three--nucleon system has been  developed by the authors in ref.[1]. 
The wave function is expanded in channels,
as in the Faddeev
technique, and for each channel,
the radial amplitude is, in turn, expanded in terms of correlated functions,
constituting the Pair correlated Hyperspherical Harmonic
(PHH) basis; such a basis  results to be well suited to carefully take into 
account the correlations induced by the large repulsive terms of the 
nucleon--nucleon (NN) potential. This technique was applied in [1] to calculate
the bound state w.f. of the triton, with the Argonne 14 (AV14) model of NN
interaction [2], with results in complete
agreement with  those given by the best available techniques [3--5].
\smallskip
It is well known that local NN pairwise potentials, as determined 
by fitting the two nucleon scattering data, do not give the correct binding 
energies of systems with A$\geq3$. If
such potentials provide a very precise fit to the data
($\chi^2$ per datum $\approx 1$), their predictions for the triton binding
energy nearly coincide [6] among themselves;
nevertheless, appreciable differences continue to exist
with the corresponding experimental values.
The lack of binding in the three--nucleon systems with
these  models, has been attributed to different possible causes, such as
relativistic corrections, non--local effects and three body interaction
(TBI) terms in the Hamiltonian. Charge symmetry breaking (CSB)
terms are also important to understand the mass difference between
$^3$H and $^3$He, but give a minor contribution to the total energy. Of
course, all these effects are not independent from each other and their
correct description is a difficult task.
\smallskip
One of the most important aims in few-nucleon theory is testing 
realistic NN potentials, not only in the bound states, but in continuum 
states too, mainly in those related to the A=3 systems. As a matter of fact, 
a series of theoretical investigations has been devoted to the study of
the p--d and n--d elastic scattering and break--up  reactions. The general 
status of the field is satisfactory [7]
and an overall agreement between theory and experiments has been 
reached. Among the possible improvements,  here we shall be interested
in a correct treatment of the Coulomb forces and in a detailed comparison
of theoretical and experimental results. To this end we have extended the 
method of ref.[1] to describe N--d scattering processes and, in this paper,
our attention will  be devoted to the study of states below the deuteron     
break--up treshold. The break--up reactions will be the object of a 
subsequent paper.

We are here particurarly interested to calculate
scattering lengths and phase shifts for the various N--d elastic
scattering channels, to be compared with the existing experimental
analysis [8]. However, such quantities are strictly related to
the ground state energy ( for example, the doublet scattering lenght and
the binding energy lie on the so--called Philips line [9]). 
As already atated,
the AV14 interaction alone, as considered in ref.[1], is not 
adequate and for this reason, we have included TBI terms in the
the Hamiltonian. These terms have been proposed
to take into account all the important two pion exchange effects,
in an approximate but quantitative way. We will consider in this paper
the so--called
Tucson--Melbourne (TM) [10] and Brazil (BR) [11] three--body potentials.
\smallskip
One important aspect of the present calculations is the extension of the
variational method to scattering states.
The Rayleigh--Ritz variational
principle has a long tradition in nuclear physics and it has been largely
used, starting from  the first attemps to investigate the bound state of
few--nucleon systems.
On the other hand, a few calculations exist for scattering 
states and, among them,
those using variational techniques, as the Kohn--Hult\'en principle,
are scarce. The reason is related to the difficulty of 
obtaining accuracies comparable with those of
bound state calculations.

In recent times, accurate variational techniques were developed to
solve the three-- [1,3] and four--nucleon [12] problems, showing a fast
convergence in the number of channels when
compared with the Faddeev techniques.
It is natural to ask whether the situation would be similar
in the case of scattering states.
This is one more motivation for the present work.
\smallskip
The paper is organized in the following way: in section 2 the construction
of the trinucleon bound state w.f. in terms of the PHH basis functions
is worked out and the resulting equations are solved with the inclusion
of TBI terms; in the next section the method is extended to describe  
N--d scattering states below the deuteron break--up threshold; the
basic equations which allow to determine the w.f and the reactance
matrix $\Re$ are obtained in the frame of the Kohn variational principle; the
numerical procedure used to solve those equations is shortly outlined
in sections 2 and 4 and, in the latter 
the results obtained for the scattering lengths
and phase shifts are given. Finally, the merits of the approach are
discussed in the last section, together with its possible extension to other
problems.
\bigskip

\noindent{{\bf 2. The three--nucleon bound state.}} 
\medskip

The three--nucleon Hamiltonian is taken to have the form

$$ H= T+\sum_{i<j}V(i,j)+\sum_{i<j<k}
      W(i,j,k)\ ,\eqno(2.1)$$
where T is the non relativistic total kinetic energy operator and the two--
and three--body interactions, $V$ and $W$,
are explicitely included. The Coulomb p--p repulsion, if it is present,
is contained in the two--body potential; moreover, the three particles are 
assumed to have the same mass ($\hbar^2/M=41.47$ MeV fm$^2$ 
throughout the paper). Let us introduce, in the center of mass (c.m.)
frame, the following set of Jacobi coordinates
($i$, $j$, $k=1$, $2$, $3$ cyclic)
     $${\bf x}_i= {} ({\bf r}_j-{\bf r}_k) , \quad
       {\bf y}_i= {2\over\sqrt 3} ({\bf r}_j+{\bf r}_k-2{\bf r}_i)\ .
         \eqno(2.2)$$\noindent
 The w.f. of the system is written as a sum of three Faddeev amplitudes, 

   $${\rm \Psi}=\psi({\bf x}_i,{\bf y}_i)+
         \psi({\bf x}_j,{\bf y}_j)+
         \psi({\bf x}_k,{\bf y}_k)\ . \eqno(2.3)$$\noindent
In the above equation, each amplitude corresponds to a total angular
momentum $J J_z$ and total
isospin $T T_z$, therefore, if we use the L--S coupling, it can be written 
in the form

$$\eqalignno{
     \psi({\bf x}_i,{\bf y}_i)=&\sum_{\alpha=1}^{N_c} \Phi_\alpha(x_i,y_i)
     \Y_\a (jk,i) &(2.4a) \cr
     \Y_\a (jk,i) =&  
     \Bigl\{\bigl[ Y_{\ell_\alpha}(\hat x_i)  Y_{L_\alpha}(\hat y_i) \bigr 
      ]_{\Lambda_\alpha} \bigl [ s_\alpha^{jk} s_\alpha^i \bigr ]_{S_\alpha}
      \Bigr \}_{J J_z} \; \bigl [ t_\alpha^{jk} t_\alpha^i \bigr ]_{T T_z}
           \ ,&(2.4b)}$$\noindent
where 
$x_i$, $y_i$ are the moduli of the Jacobi coordinates. Each $\alpha$--
channel is specified by the  angular momenta  $\ell_\alpha$, 
$L_\alpha$ coupled to give $\Lambda_\alpha$, and by the spin
(isospin) $s_\alpha^{jk}$ ($t_\alpha^{jk}$) and $s_\alpha^i$
($t_\alpha^i$) of the pair  $j$, $k$ and the third particle $i$,
coupled to give $S_\alpha$ ($T$). The number $N_c$ of channels taken into 
account to construct the w.f. can be increased until 
convergence is reached. 
The antisymmetrization of the  w.f. $\Psi$ requires 
$\ell_\a  + s^{jk}_\a + t^{jk}_\a $ to be odd; in addition  
$\ell_\a + L_\a$ must be even for 
positive parity states and odd for the negative ones. 

The ground state
of the three--nucleon system has positive parity and $J=1/2$, $T=1/2$ 
(the inclusion of the Coulomb potential gives a small $T=3/2$ contribution,
disregarded in the present work, as well as other
charge symmetry breaking terms in the Hamiltonian).
The channels allowed by such
conditions are easily obtained and ordered of increasing angular momenta 
values; for example, they are 10, 18 and 26 for
$\ell_\a +L_\a \leq 2$,  $\ell_\a +L_\a \leq 4$ and
$\ell_\a +L_\a \leq 6$, respectively.

Let us introduce,
in place of the coordinates $x_i$, $y_i$,
the hyperspherical coordinates 
defined by:

$$  x_i =\rho \cos\phi_i\ ,\quad y_i =\rho \sin\phi_i \ ,\eqno(2.5)$$
where $\rho$ is the hyperradius.
The radial dependence of each $\a$--amplitude in the w.f. (2.4$a$) is now 
expanded in terms of the PHH basis functions in the following way:
  $$\Phi_\alpha(x_i,y_i)
     =\rho^{\ell_\a + L_\a}f_\a (x_i)
     \Bigl[ \sum_{K=K_0}^{K_\a} u^\alpha_K(\rho)\ {} 
     ^{(2)}P^{\ell_\a ,L_\a}_K(\phi_i)\Bigr] 
           \ ,\eqno(2.6)$$\noindent
where the hyperspherical polynomials are given by [13]
$$^{(2)}P_K^{\ell_\a ,L_\a}(\phi_i)=N_n^{\ell_\a ,L_\a }
        (\sin\phi_i)^{L_\a}(\cos\phi_i)^{\ell_\a}
        P_n^{L_\a +1/2,\ell_\a +1/2}(\cos{2\phi_i})\ ,\eqno(2.7)$$
$N_n^{\ell_\a ,L_\a}$ is a normalization factor and $P_n^{\a ,\beta}$
is a Jacobi polynomial. The grand orbital quantum number is given by
$K=\ell_\a + L_\a +2n\,$, with $n$ a non--negative integer. 
In eq.(2.6) $K_0=\ell_\a + L_\a $ is the 
minimum grand orbital quantum number and $K_\a$ is the maximum selected value,
so that the number of basis functions per channel is 
$$ M_\a = (K_\a - K_0)/2 + 1\, , \eqno(2.8)$$
corresponding to
the maximum value of the index $n$ plus one. When $\alpha$ 
goes to infinity, the expansion basis used in eq.(2.6) is obviously complete.
\smallskip

If the functions $f_\a (x_i)$ in eq.(2.6) are taken equal to one, 
the standard (uncorrelated) HH expansion is recovered. Such an expansion is 
well suited to describe the structure of the system in the case of soft 
interparticle potentials, where a rather small number of basis functions 
is sufficient to reproduce the w.f. within a reasonable accuracy [14]. 
However, for potentials containing a strong repulsion at small 
distances, the w.f. must be accurately determined for small interparticle 
separation values and correspondingly the rate of convergence of the HH 
expansion results to 
be very slow [15]. The role of the correlation function $f_\a (x)$ in 
eq.(2.6) is therefore to fasten the convergence of the expansion by improving 
the description of the system when a pair of particles are close to each other. 
A simple procedure to determine the correlation functions is 
the one  outlined in [1],where the functions $f_\a (r)$ are taken as
the  solutions of the following
zero--energy Schroedinger equations

$$ \sum_{\beta'}[ T_\bb (r) + V_\bb (r) + \lambda_\bb (r) ]
    f_{\beta'}(r)=0. \eqno(2.9) $$
 $T_\bb$ and $V_\bb$ are the kinetic and the potential energy operators, 
$$\eqalignno{
  T_\bb =&-\htm\left[{\partial^2\over \partial r^2} + {2\over r}{\partial\over
         \partial r} - {\ell_\b (\ell_\b +1)\over r^2}\right] \delta_\bb\, , 
         &\cr &&(2.10)\cr
  V_\bb =& <\ell_\b\; s^{jk}_\bp\; t^{jk}_\bp\; | V(jk)|
            \ell_\b\; s^{jk}_\b \; t^{jk}_\b \; >\, , &\cr}$$
and  $V(jk)$ is the NN potential; the term $\lambda_\bb (r)$ in (2.9) 
is chosen of the simple form [1,16]

$$ \lambda_\bb (r) = \lambda^0_\b \exp(-\gamma r) \delta_\bb \eqno(2.11)$$
and its role is to allow the function $f_\b (r)$ to satisfy an 
appropriate healing condition. This can be achieved by taking $\gamma$ as a 
trial parameter, whose precise value is not important (the choice
$\gamma\approx .5\,$fm$^{-1}$ is adequate [16]),
and conveniently fixing the depth $\lambda^0_\b $.
As an example, in the case of an uncoupled channel we can require that
$$ f_\b (r)=1,\quad{\rm when}\quad r>R, \eqno(2.12)$$
where $R$ is large with respect to the range of the potential $V_\bb (r)$. 
For coupled channels, the condition (2.12) is satisfied by the function 
associated with the lower angular momentum value, while the other function 
goes to zero when r becomes large. 
\smallskip  
The remaining problem is the determination of the  hyperradial
functions $u^\alpha_K(\rho)$ contained in 
eqs.(2.6); to this aim we will use the Rayleight--Ritz principle,
requiring that the following condition to be satisfied
$$ <\delta_u\Psi|H-E|\Psi>=0\, , \eqno(2.13)$$
where $\delta_u\Psi$ represents the change in the w.f. caused by an 
infinitesimal 
variation of the functions $u^{\a}_K (\rho)$. From the latter equation, it 
follows that
$$ \rho^{\ell_\a + L_\a } \sum_i <f_\a(x_i)\; ^{(2)}P_K^{\ell_\a ,L_\a}(\phi_i)
    \Y_\a (jk,i)|H-E|\Psi>_{\O} =0\, , \eqno(2.14)$$
where the subscript $\O$ indicates that the integration over the hyperangles 
$\phi_i$ and the angles $\hat x_i, \hat y_i$ must be performed, and  
$ d\O = \sin^2\!\phi_i \cos^2\!\phi_i\, d\phi_i d\hat x_i d\hat y_i $. 
From  eq(2.14), after the evaluation of the spin--isospin traces and
the angular integration, one obtains a set 
of second--order differential equations for the functions 
$u^{\alpha'}_{K'}(\r )$
which can be written in the form [1]
$$\sum_{\alpha',K'}
       \Bigl[ A^\aa_\kk (\r ){d^2\over d\r^2}+ B^\aa_\kk (\r ){d\over d\r}
            + C^\aa_\kk (\r )+{m\over\hbar^2} E\; N^\aa_\kk (\r )\Bigr ]  
              u^{\alpha'}_{K'}(\r )= 0\ ,    \eqno(2.15)$$\noindent
with $\alpha' =1,\ldots,N_c$ and $K^\prime=K^\prime_0,\ldots,K^\prime_\ap$.
A numerical technique for  solving the set of eqs.(2.15) 
has been outlined in ref.[1]. An important point is to reduce as
much as possible the number $N$
of grid points $[\rho_1,\ldots,\rho_N]$ in the hyperradius, without losing 
accuracy in the solution. Due to the nature of the problem, the coefficients
$X^\aa_\kk\ (X=A,B,C,N)$, and correspondingly the solution, strongly vary only
for small values of $\rho$; for this reason, a grid of the type
$\rho_{k+1}-\rho_k=\chi(\rho_k-\rho_{k-1})$, where the step--length is
increased by a constant factor,
results to  be a convenient choice. As a consequence, a new variable $w$ is
introduced with equally spaced grid values $w_k$ and the corresponding
$\rho_k$ values can be obtained by the relation
$$ w={h\over \ln\chi}\ln[1+{\rho\over h}(\chi-1)]\eqno(2.16)$$
with $h=\rho_2-\rho_1$. For potentials without 
hard-core repulsion, as the ones considered here, we have $\rho_1=w_1=0$,
and the last grid point $w_N$ can be chosen
so that $\rho_N=\rho_{max}$ is about
$20 \div 25$ fm for bound state calculations.
The set of eqs.(2.15) can be  re--written in terms of the  variable $w$ and
the corresponding generalized eigenvalue problem then solved by standard
procedures.

Another useful possibility is to introduce a new variable to
map the infinite range of the hyperradius $[0,\infty]$ into the finite
interval $[0,1]$. The new variable is then transformed 
into the variable $w$ which 
is used with a 
constant step grid. This is achieved by the following two
transformations:

$$\eqalignno{
    z&=1-\exp(-\nu\rho) &\cr
    w&={h\over \ln\chi}\ln[1+{z\over h}(\chi-1) ]\ ,&(2.17)\cr}$$
where the parameters $\nu$ and $\chi$ are chosen in such a way to get the most 
convenient set of $\rho_k$ values. 
The main difference between the trasformations (2.16) and (2.17) lies in the
fact that, when solving the set of differential equations,
a boundary condition mut be imposed at $\rho=\rho_{max}$ in the first case,
whereas
in the second case, the final point in the  $w$--grid
corresponds to $\rho=\infty$, and for such a point the usual boundary
condition is that the function to be determined is zero. In the 
calculation of the A=3 bound state w.f., both the transformations (2.16) and
(2.17) produce almost exactly the same results. However, in the study of 
scattering states, the asymptotic conditions to be satisfied are more easily 
fulfilled
by the transformation (2.17), which will be therefore the one adopted in 
the next section.
\smallskip
   The ground state of the triton has been studied in ref.[1] by
means of the PHH approach with the AV14 potential. The system of
equations (2.15) has been solved
by including up to 12 channels in the expansion (2.6); for the sake of
completeness, we have extended the calculations to include 
all the channels with $\ell_\a + L_\a \leq 4$ and the results are
displayed in table 1. By inspection of the table, one can see that
the PHH expansion is rapidly convergent to results which are in
complete agreement with those obtained in ref.[3], where a
different technique for constructing the w.f. has been used. 
These
circumstances should eliminate any doubt about a still
missing small contribution to the binding energy, due to the use of variational
bases in both the approaches. 
\smallskip
We have also studied the structure of the A=3 systems when 
three--body forces are included in the Hamiltonian of the system. Therefore,
we have implemented the nucleon interaction by adding
to the AV14 potential both the
Tucson--Melbourne (TM) and the Brazil (BR) three--body potentials, which
have been used by different groups [3,17,18]
to study the three--nucleon bound state.
In these models the
numerical choice of the  $\pi$N form factor cutoff $\Lambda$ appears to be
critical, the
choice $\Lambda=5.8\mu$ 
($\mu$ is the pion mass) checked in ref.[17],
leads to an overbinding of the trinucleon system. Of course, it is
possible to consider also other TBI effects, as for example
those due to $\rho$--meson exchanges [19]. 
In order to compare with the mentioned calculations, as a
first choice we have adopted the same $\Lambda$ value of ref.[17].
The results obtained by
including up to 18 channels are shown in  table 2, together with the 
estimates of the Los Alamos group corresponding to a 34 channels 
configuration space Faddeev (CSF) 
calculation [17]. The agreement between the two approaches is very 
satisfactory. Successively, the value of the cutoff 
parameter $\Lambda$ has been changed
in order to provide a triton binding energy value close to 
the experimental one. The results are listed in table 4 for the AV14 + TM
and AV14 + BR models and correspond to the values
$\Lambda=5.13\mu$ and $\Lambda=4.99\mu$, respectively.
As mentioned in the introduction, 
it is necessary that the 
adopted potential models give a satisfactory description of the A=3 bound
states, due to the correlation between low energy
scattering observables and ground state binding energy,
and for this reason the modified values of the $\Lambda$ parameter
previously given will be used in the next section.
\smallskip
Finally, the  $^3$He ground state could easily be calculated  
with the same TBI terms
and with the Coulomb potential. However, it is well known that the
Coulomb interaction alone is not able to completely explain the mass difference
between $^3$H and $^3$He and 
other CSB terms must be added to the Hamiltonian [18,19].
This problem will 
no longer be discussed here. 
\bigskip

\noindent{\bf 3. N-d scattering below the break--up threshold}\medskip
 
The progress in the experimental and theoretical study of
N--d interaction processes during the last 
decade has been noticeable. In particular, N--d elastic 
scattering and break-up cross sections data are now available over a large  
range of energies. Accurate results have been also obtained for
polarized nucleon and deuteron targets [8]. The theoretical analyses
have been done using a variety 
of techniques and NN potentials [20,21],
including three-body forces too [20,22]. 
Moreover, the treatment of the Coulomb interaction for the pd system has 
been improved in a satisfactory way [23]. Nevertheless, the interest in 
this field remains noticeable, first of all because of the possibility
of testing models and techniques by
comparing the calculated and experimental values for a number of important 
observables. For such
a reason, it is evident the importance of increasing
the accuracy both in the experimental and theoretical studies. 
\smallskip
The 
variational approach based on the use of PHH correlated functions can be 
extended to investigate scattering states and in this section the 
application to the N--d scattering below the break--up threshold is discussed. 
Following the pioneering work of Delves [24] for realistic NN 
interactions,
the wave function for a N--d scattering state will be written as

$$ \Psi = \Psi_{C} + \Psi_{A}\ .\eqno(3.1)$$

The first term $\Psi_C$ must be sufficiently flexible to guarantee a detailed 
description of the ``core'' of the system, when the
particles are close to each other and the mutual interaction is large;
$\Psi_C$
goes to zero when the nucleon--deuteron distance $ r_{Nd} $ 
increases. As in the previous section, $\Psi_C$ is the sum of three
Faddeev amplitudes which are in turn expanded
in terms of the PHH basis functions.
The second term $\Psi_A$ of eq.(3.1) has to describe the asymptotic 
configurations of the system, for large $r_{Nd}$ values, where the 
nuclear N--d interaction is negligible. In the asymptotic region the w.f.
$\Psi$ reduces to  $\Psi_A$, which therefore must be the appropriate
asymptotic solution of the 
Schroedinger equation. $\Psi_A$  can also be decomposed in
three Faddeev amplitudes and each one of these is written as a
linear combination of the following functions

$$ \eqalignno{
   \Omega^\lambda_{LSJ}(\x_i,\y_i) =&\sum_{l_\a=0,2} w_{l_\a}(x_i)
       {\cal R}^\lambda_L (y_i) \times&\cr
       &\left\{\left[ [\yx s_\a^{jk}]_1 \otimes s^i \right]_S 
       \otimes \yyl \right\}_{JJ_z}
       [t_\a^{jk}t^i]_{TT_z}\ .&(3.2)\cr}$$
In this equation  $w_{l_\a}(x_i)$ is the deuteron wave function
component in the waves with $l_\a =0,2$; $L$ is the relative angular
momentum of the deuteron and the 
incident nucleon,  $S$ is the spin obtained by coupling the spin 1 of the
deuteron to the spin 1/2 of the incident nucleon. Therefore, an
asymptotic state will be labelled as $ ^{(2S+1)}L_J$ and the
corresponding phase shift as $\delta_{LSJ}$.
\smallskip
The functions ${\cal R}^\lambda_L (y_i)$ of eq.(3.1) can be taken as
the regular ($R$) and irregular ($I$) radial solutions
of the two--body (N--d) Schroedinger equation without nuclear interaction.
The regular solution, denoted as $ {\cal I}_L(y_i)$, can be written in 
the form

$$ {\cal I}_L (y) = { F_L(\eta,\zeta)\over (2L+1)k^L\zeta C_L(\eta)}
   \ ,\eqno(3.3)$$
where $\eta = 2Me^2/3\hbar^2k$ and $\zeta=kr_{Nd}$ are
the usual Coulomb parameters,
$k$ is related to the center of mass energy $ E_{c.m.}=3\hbar^2k^2/4M $
and $y=(\sqrt3/2)r_{Nd}$.
The regular Coulomb function $F_L$ and the factors $C_L$ are
defined in the standard way and the non--Coulomb case is obtained in the limit
$ e^2 \rightarrow 0$ (see also ref.[25]). The irregular
solution, denoted as $ {\cal K}_L (y_i) $, has the form

$$ {\cal K}_L (y)= (2L+1) k^{L+1} C_L(\eta) {G_L(\eta,\zeta) 
   \over \zeta } \eqno(3.4)$$
where $ G_L(\eta,\zeta)$ is the irregular Coulomb function. The function 
${\cal R}^\lambda_L (y_i)$ of eq.(3.1) is taken equal to
${\cal I}_L(y)$ for $\lambda\equiv R$, whereas for $\lambda\equiv I$
it does not coincide with
${\cal K}_L(y)$, since this would introduce a singular behaviour at
$r_{Nd}=0$, 
which should be corrected by the $\Psi_C$ term in eq.(3.1). In 
order to avoid it, the function $G_L(\eta,\zeta)$ in 
eq.(3.4) has been replaced by ${\widetilde G}_L(\eta,\zeta)$, which differs 
from the previous one by a regularizing factor. Of course, the detailed form
of such a factor is not of particular relevance, and in conclusion the 
following simple form will be used

$$ {\widetilde G}_L(\eta,\zeta)= (1-\exp^{-\xi r_{Nd}})^{L+1} G_L(\eta,\zeta)
    \ .  \eqno(3.5)$$
The trial parameter $\xi$ is determined  by requiring that
${\widetilde G}_L$ tends to $G_L$ smoothly and the value
$\xi= 0.25 \fm^{-1}$ is found to be adequate.
With the above definitions, the $i$--th Faddeev amplitude for the
asymptotic wave function is written as

$$\Omega_{LSJ}(\x_i,\y_i) = \Omega^R_{LSJ}(\x_i,\y_i) + \sum_{L'S'} 
         {}^J\R^{SS'}_{LL'}\Omega^I_{L'S'J}(\x_i,\y_i)\ , \eqno(3.6)$$
where the matrix elements $^J\R^{SS'}_{LL'}$ give the relative weight
between the regular and the irregular components. They are
closely related to the corresponding 
reactance matrix ($\Re$--matrix) elements:

$$ ^JR^{SS'}_{LL'}= (2L+1)(2L'+1) k^{L+L'+1}
                      C_LC_{L'}\ {}^J\R^{SS'}_{LL'}\ .\eqno(3.7)$$
By definition, the eigenvalues of the $\Re$--matrix are
$\tan\delta_{LSJ}$.

The internal part $\Psi_C$ of the w.f. (3.1)
is decomposed in the three Faddeev amplitudes and each one
is expanded in terms of the PHH basis, as it was done for the bound state

  $$\eqalignno{
     \psi_C({\bf x}_i,{\bf y}_i)=&\sum_{\alpha=1}^{N_c} \Phi^C_\alpha(x_i,y_i)
     \Y_\a (jk,i) &(3.8a) \cr
     \Phi^C_\alpha(x_i,y_i)=&\rho^{\ell_\a + L_\a}f_\a (x_i)
     \Bigl[ \sum_{K=K_0}^{K_\a} u^\alpha_K(\rho)\ {} 
     ^{(2)}P^{\ell_\a ,L_\a}_K(\phi_i)\Bigr] 
           \ ,&(3.8b)\cr}$$\noindent
with the conditions $u^\alpha_K(\rho) \rightarrow 0 $ when
$\rho \rightarrow \infty $.

The form of the $\Psi_A$ component is given by eqs.(3.2) and (3.6)
also at small interparticle distances, therefore the
internal function $\Psi_C$ must properly correct,
such a $\Psi _A$ behaviour, in that region.
For this reason, the partial wave decomposition of $\Psi_C$
(which is truncated when doing calculations)
must include, first of all,
those channels that are present in the asymptotic state (open channels),
but the summation must be extended also to all
the other important $\alpha$--channels compatible with the state 
to be described.
In conclusion, the total wave function corresponding to an
asymptotic state ${}^{(2S+1)}L_J$ will be written as

$$ \eqalignno{\P_{LSJ} =&\sum_{i=1,3}\left[
     \Phi_C(\x_i,\y_i) + \Omega_{LSJ}(\x_i,\y_i)\right] &\cr
   =&\sum_{i=1,3}\left[
     \Phi_C(\x_i,\y_i) + \Omega^R_{LSJ}(\x_i,\y_i) +\sum_{L'S'}
    {}^J\R^{SS'}_{LL'}\Omega^I_{L'S'J}(\x_i,\y_i)\right] \ . &(3.9)\cr}$$

The Hamiltonian connects states with the same parity and total angular
momentum $J$, but with $\Delta L=0,1,2$. As a consequence, the 
$\Re$--matrix for $J=1/2$ is a $2\times2$ matrix and
it is a $3\times3$ matrix in all the remaining cases. So, for a 
given $J$ two or three independent
functions (3.9) can be built up by
different combinations of $L$ and $S$.

The quantities to be determined in the wave functions (3.9) are the hyperradial 
functions (see eq.(3.8$b$)) and the matrix elements $^J\R^{SS'}_{JJ'}$.
To this aim we will use the Kohn variational principle.
This variational principle establishes that, for scattering states, the
$\Re$--matrix elements, considered as functionals of the wave function,
must be stationary [23] with respect to 
variations of all the trial parameters. Explicitely, these
functionals are given by:

$$\eqalignno{
  [{}^J\R^{SS'}_{LL'}]&={}^J\R^{SS'}_{LL'}-<\P_{L'S'J}|{\cal L}|\P_{LSJ}>
                     &(3.10)\cr
      {\cal L}&={M\over 2\sqrt3\hbar^2}(H-E)\ ,&(3.11)\cr}$$
where $\ ^J\R^{SS'}_{LL'}$ are the trial parameters of eq.(3.9). 
With the above definition of the operator ${\cal L}$ and using
eqs.(3.2--4), we fix the normalizations of the asymptotic states as

$$  <\Omega^R_{LSJ}|{\cal L}|\Omega^I_{LSJ}>
   -<\Omega^I_{LSJ}|{\cal L}|\Omega^R_{LSJ}> = 1\ .\eqno(3.12)$$

\smallskip
The variation of the diagonal functionals with respect to the hyperradial
functions $u^\alpha_K(\rho)$ is first performed. From the condition

$$ \delta_u [{}^J\R^{SS}_{LL} ] = 
       <\delta_u\P_{LSJ}|{\cal L}|\P_{LSJ}>=0,  \eqno(3.13)$$
and using the same procedure as in the case of the
Rayleigh--Ritz variational principle for to the bound state,
an inhomogeneous set of second order differential equations is obtained:

$$\sum_{\alpha',K'}
       \Bigl[ A^\aa_\kk (\r ){d^2\over d\r^2}+ B^\aa_\kk (\r ){d\over d\r}
            + C^\aa_\kk (\r )+{m\over\hbar^2} E\; N^\aa_\kk (\r )\Bigr ]  
              u^{\alpha'}_{K'}(\r)= D^\lambda_{\alpha K}(\rho)
  \ .    \eqno(3.14)$$\noindent
The  coefficients $A,B,C,N$ have the same expression as in the
bound state case, and the inhomogeneous term is given by

$$ D^\lambda_{\alpha K}(\rho)= \rho^{l_\a + L_\a}
   \sum_{ii'}<f_\a(x_i)\,{}^{(2)}P^{l_\a L_\a}_K(\phi_i)\Y_\a (jk,i)|{\cal L}
   |\Omega^\lambda_{LSJ}(\x_{i'},\y_{i'})>_\O \ ,\eqno(3.15)$$
where, the subscript $LSJ$ has been omitted for simplicity.
For each asymptotic state $^{(2S+1)}L_J$  two different inhomogeneous terms
can be constructed in correspondence to the
asymptotic $\Omega^\lambda_{LSJ}$ function with
$\lambda\equiv R$ or $I$. Correspondingly, two different sets of
hyperradial functions are obtained by solving 
the system of eqs.(3.14). 
\smallskip
Then, in order to get the optimum choice for the
matrix elements $\ ^J\R^{SS''}_{LL''}$,
the diagonal functionals (3.10) are varied 
with respect to them. This leads to the following set of
algebraic equations

$$ \sum_{L'',S''}{}^J\R^{SS''}_{LL''} X^{S'S''}_{L'L''}
    = Y^{SS'}_{LL'} \ ,\eqno(3.16)$$
with the coefficients $X$ and $Y$ defined as

$$\eqalignno{
  X^{S'S''}_{L'L''}&= <\O^I_{S'L'J}+\Psi^I_{S'L'J}|{\cal L}
                       |\O^I_{S''L''J}>&,\cr
                   &&(3.17)\cr
  Y^{SS'}_{LL'}&= -<\O^R_{SLJ}+\Psi^R_{SLJ}|{\cal L}
                       |\O^I_{S'L'J}>&\cr}$$
In the latter equations, $\Psi^\lambda_{LSJ}$ indicates the internal part
of the wave function constructed with one of the two 
solutions of eqs.(3.14), as previously obtained.

It is worth to notice that the variation of the diagonal functionals 
gives different coupled equations which must be satisfied
by the off--diagonal elements of the
reactance matrix [26]. However, the corresponding solutions
are a first order variational
estimates and the last term in eq.(3.10) is, in general, not zero, 
as it would happen in the case of
the exact wave function. A second
order estimate for the $\Re$--matrix elements can be obtained by substituting,
in the second member of equation (3.10), the first order results.
As the reactance matrix is symmetric, the method should provide 
${}^J\R^{SS'}_{LL'} = {}^J\R^{S'S}_{L'L}$. The degree of violation of this
condition give useful information about the accuracy of the solution.
\bigskip

\noindent{\bf 4. Numerical results}\medskip

The numerical technique to solve the inhomogeneous linear equations
system (3.14) deserves some attention. The relevant point is that the
scattering state w.f. (3.1) must carefully describe both the regions of 
small and medium--large interparticle distances, the asymptotic behaviour 
having been properly taken into account throught the $\Psi_A$ component. 
For this reason, it is convenient to use in place of $\rho$ the variable 
$w$, defined by the transformation (2.17). By using this new variable, and 
after replacing the differential operators by finite differences, the 
system (3.14) is transformed into a set of algebraic linear equations 
which can be solved by means of standard methods. We have used
$N=40 \div 50$ $w$--grid points have been used, the last point $(w_N=1)$ 
corresponds to $\rho=\infty$ and the penultimate $w$ value corresponds
to $\rho_{N-1}\approx 65.0\,$fm. 
\smallskip
The first case considered is a zero energy
scattering process. In this case,
due to centrifugal barrier effects, the reactance matrix is diagonal
and two physical states with $L=0$ can be constructed:
the $J=1/2$ (doublet) and the $J=3/2$ (quartet) states. The
corresponding scattering lenghts are defined as

$$  {}^{(2J+1)}a_{N-d} = -\lim_{k\to0}
                      {{}^J\R^{JJ}_{00}\over k} \ . \eqno(4.1)$$

The first NN interaction considered is the semi--realistic
Malfliet and Tjon [27] potential (I--III) acting only
in the $s$--wave. For this potential
the `` exact'' results of the CSF method [28] are available, therefore
a meaningful comparison can be done. Our results are
${}^2a_{n-d}=0.702\,$fm and ${}^2a_{p-d}=0.003\,$fm for the doublet state
and ${}^4a_{n-d}=6.442\,$fm and ${}^4a_{p-d}=13.96\,$fm
for the quartet one. These values are extremely close to
those given by the CSF method. 

Then we have considered the realistic 
AV14 and AV14+BR interactions, with the TBI cutoff parameter 
$\Lambda=5.8\mu$. The calculated
doublet and quartet scattering lengths are reported
in table 4 as a function of the number $N_c$ of channels included
in the expansion of the w.f., together with the 34--channels CSF
results of ref.[20]. The agreement between the two methods is quite
satisfactory, and there are only small differences due to the 
contributions from higher channels not present 
the CSF approach.

If the value of the cutoff parameter is taken $\Lambda=4.99\mu$,
so as to reproduce the correct triton binding energy, the resulting
scattering lengths are again very close to the CSF ones given in
ref.[20].  Concerning the comparison between the theoretical
predictions and experimental data,
here we only mention that for the n--d case the agreement is acceptable,
whereas, as it is discussed in ref.[25], one
must be very carefull, for the p--d system, when extrapolating to zero energy
the experimental phase shifts obtained at not sufficiently low energies.

The numerical analysis has then been extended to elastic Nd scattering
states below the deuteron break--up treshold. For scattering energy
different from zero, the reactance
matrix, with the exception of states having J=1/2, is a
symmetric $3\times 3$ matrix. Its six independent
parameters are the three eigenvalues ($\tan{}\delta_{LSJ}$) and the three
mixing parameters which are here introduced by adopting
the formalism of Seyler [29]. As it was mentioned in section 3,
the solution obtained by means of eqs.(3.16) gives different
first--order estimates for the off--diagonals elements. However,
the second--order estimates, calculated with the help of eq.(3.10)
should verify the symmetric condition with an improved  accuracy.
In order to stress this aspect of the problem, the first--
and second--order estimates for the $\Re$--matrix elements, 
calculated with the AV14 potential, are shown in table 5
in the case $J=1/2^+$ and a neutron incident energy of 3.0 MeV,
The first four rows
of the table correspond to a calculation with a number of channels
$N_c=8$ and $M=3$ hyperradial functions per
channel. For the successive four rows $N_c=8$ and $M=6$
have been used and the last four rows are obtained with $N_c=10$ and $M=6$.
By inspection of the table, it can be seen that
the second order variational calculation accurately
provides a symmetric reactance matrix and the correction given by the
mean value of the operator ${\cal L}$ goes to zero as the number of
hyperradial functions and channels increase.
Analogous rates of convergence have been found for all the states
studied in the present work.
\smallskip
Accurate experimantal data and systematic phase shifts analyses exist
for the pd system [8], for incident proton energy values
$E_N$=1.0, 2.0, 3.0 MeV. We have 
calculated the scattering states with the above energy values
having in mind the two following motivations. First a detailed 
comparison with the experimental results using realistic interactions
(including also  three-body terms) and an adequate treatment of
the Coulomb repulsion,
is of interest to judge the merits of the 
theoretical underlying model; second, the calculated $\Re$--matrix
elements may be useful in the case of future
phase shifts analyses of accurate experimental data.
\smallskip
Coming back to our calculation, the phase 
shifts with $L>2$ can be 
disregarded at the energies here considered, 
so only twenty independent parameters are contained
in the cross section, i.e. thirteen phase shifts and seven
mixing parameters. The variational estimates of these
twenty parameters are displayed in table 6
at three different values of the scattering energy. For sake of completeness,
the system nd has also been studied and the results are
reported in the table. 
The contribution from the TBI terms 
is unimportant in all the states, with the exception of the J=1/2$^+$, 
a circumstance already noticed in ref.[22]. The values listed in table 6 
correspond to the AV14 potential, but when the AV14+BR 
predictions differ appreciably from the AV14 ones,
they are reported in 
parentheses in the table.
Concerning the number of channels, the value
$N_c=14$ for $J=1/2^+$ and $N_c=18$ for $J=3/2^+$ have been used.
For the other $J$ values $N_c$ has been fixed by including all
open channels (giving the major contribution) and a few selected
close channels which give a minor but appreciable
contribution; in any case the total number of channels does never exceed
$N_c=18$. The
number of hyperradial functions per channel can vary in the range
$M=3\div6$, with enough flexibility in the radial functions
to get convergence for all the calculated 
matrix elements listed.
\bigskip

\noindent{{\bf 5. Discussion and conclusions}}
\medskip

The study of the scattering processes involving a few strongly interacting 
particles is a field of large interest, but only in recent times it has been 
possible to performe accurate numerical calculations of the important 
related quantities. An emblematic situation is that one involving three 
nucleons and, in particular, the N--d interaction processes. For this
system, the 
main progress has been realized by the Faddeev method,
both in momentum and coordinate representations. A 
delicate aspect of such an approach is connected with the treatment of the 
pp Coulomb interaction. As it is well known, the nuclear and Coulomb 
potentials are expanded in channels with increasing angular momenta 
and the expansion is enlarged until a satisfactory convergenge for 
the w.f. of the system and the (part of) potential taken into account
is found.
However, the expansion in partial waves of the 
Coulomb interaction 
is slowly convergent and the problem of the missing contribution must be 
carefully investigated. 
\smallskip
On the other side, the merit of a
variational treatment is that even
if the w.f. is expandend in channels, the 
interaction is fully taken into account as no expansion of 
the potentials is performed. Of course, the variational tecnique 
must be devised with particular attention in order to
reproduce all the relevant 
details of the investigated structure. This might appear as easy 
to be satisfy, since the number of trial parameters can be increased 
and the computational facilities available at present allow to handle 
quite easily the
corresponding numerical problem. However, this is
the situation 
only for rather simple homework interactions and the variational analyses of 
the bound and scattering states of the three--nucleon system,
with realistic interactions, have 
encountered severe difficulties. One possible solution is to employ 
adequate expansion basis for the radial dependence of each 
channel in the w.f. expansion. Since the NN interaction produces strong 
nucleon--nucleon correlations, it appears to be convenient to introduce 
sets of basis 
functions which in somehow take care of these correlations. A choice,
investigated with success in the bound state of the A=3 systems,
is a correlad harmonic oscillator (CHO) basis [30.
However, 
in order to reproduce an exponential type behaviour characteristic 
of the large interparticle distances of the 
bound state w.f. of nuclear systems ,
a great number of HO functions is 
necessary. With respect to this, the use of correlated hyperspherical harmonic 
basis appears more convenient, and the results obtained in ref.[1] for the 
bound state and, in the present paper, for elastic scattering states on three 
nucleons are very satisfactory. 
\smallskip
There are a few aspects of interest in the results presented throughout the
paper which we wish to point out. First, there are no difficulties in
treating the Coulomb interaction for A=3 scattering states, even above the
deuteron break--up threshold. The inclusion of TBI terms in the nuclear 
interaction does not cause additional problems and it is not necessary
to increase the number of the w.f. channels. Again the reason lies in the
fact that no expansion in partial waves of the TBI is required. 
Concerning the
numerical studies presented for the N--d elastic scattering processes,
we have considered one of the available so--called realistic interactions,
i.e. the AV14 potential with the Brazil three--body interaction, with
a value of the $\Lambda$ parameter chosen
to produce a correct value of the triton binding energy. However, the extension
of the variational method based on PHH correlated functions to other forms
of local or non local interactions, as for example the Bonn potential,
does not present difficulties and could be useful for precise testing of 
the adopted model. Indeed, the theoretical analyses allow for accurate
evaluations of the phase shifts and mixing parameters in the various channels.
Moreover, since there is an overall satisfactory agreement between theoretical
and experimantal data, future improvements in the accuracy of the experimental
data would certainly be of interest.
\smallskip
We would notice finally that the extension of the proposed variational
method to scattering states for systems with larger number of particles,
in particular A=4, can be easily performed but the corresponding numerical
effort strongly increases. Interesting numerical results
can, nevetheless, be again obtained.
\vfill\eject

\noindent{ {\bf References} }\bigskip
\item{1)} A.Kievsky, M.Viviani and S.Rosati, Nucl.Phys. {\bf A551} (1993) 241
\item{2)} R.B.Wiringa, R.A.Smith and T.A.Ainsworth, Phys.Rev. {\bf C29}
          (1984) 1207
\item{3)} H.Kameyama, M.Kamimura and Y.Fukushima, Phys.Rev. {\bf C40}
          (1989) 974; H.Kameyama, M.Kamimura and Y.Fukushima, Nucl.Phys.
          {\bf A508} (1990) 17c
\item{4)} C. R. Chen, G. L. Payne, J. L. Friar and B. F. Gibson, 
          Phys.Rev.{\bf C31} (1985) 266;
          J. L. Friar, B. F. Gibson and G. L. Payne,
          Phys.Rev.{\bf C36} (1987) 1138
\item{5)} S. Ishikawa, T. Sasakawa, T. Sawada and T. Ueda, 
          Phys.Rev.Lett.{\bf 53} (1984) 1877;  
          T. Sasakawa and S. Ishikawa, Few-Body Syst. {\bf 1} (1986) 3
\item{6)} J.L.Friar, G.L.Payne, V.G.J.Stoks and J.J.de Swart,
          Phys.Lett. {\bf B311} (1993) 4 
\item{7)} W.Plessas, Few--Body Syst.Suppl. {\bf 6} (1992) 265
\item{8)} P.A.Schmelzbach, W.Gr\"uebler and R.E.White,
          Nucl.Phys. {\bf A197} (1972) 273;
          J.Arvieux,
          Nucl.Phys. {\bf A221} (1974) 253;
          E.Huttel, W.Arnold, H.Baumgart, H.Berg and G.Clausnitzer,
          Nucl.Phys. {\bf A406} (1983) 443 
\item{9)} A.C.Philips, Rep.Prog.Phys. {\bf 40} (1977) 905;
          Phys.Lett. {\bf B28} (1969) 378
\item{10)}S.A. Coon, M.D. Scadron, P.C. McNamee, B.R. Barrett, D.W.E. Blatt
          and B.H.J. McKellar, Nucl.Phys.{\bf A317}, (1979) 242;
          S.A. Coon and W. Gl\"ockle, Phys. Rev. {\bf C23} (1981) 1790.
\item{11)}H.T. Coelho, T.K. Das and M.R. Robilotta, Phys. Rev. {\bf C28},
          (1983) 1812.
\item{12)}M.Viviani, A.Kievsky and S.Rosati to be published
\item{13)}M. Fabre de la Ripelle,
          Ann.Phys(N.Y.) {\bf 147} (1983) 281
\item{14)}G. Erens, J. L. Visschers and R. van Wageningen,
          Ann.Phys.(N.Y.){\bf 67} (1971) 461;
          M. Beiner and M. Fabre de la Ripelle, 
          Lett. Nuovo Cim.{\bf 1} (1971) 584
\item{15)}J. Bruinsma and R. van Wageningen
          Phys.Lett.{\bf 44B} (1973) 221:
          J. L. Ballot and M. Fabre de la Ripelle,
          Ann.Phys.(N.Y){\bf 127} (1980) 62
\item{16)}S. Rosati, M. Viviani and A. Kievsky,
          Few--Body Syst.{\bf 9} (1990) 1
\item{17)}C. R. Chen, G. L. Payne, J. L. Friar and B. F. Gibson, 
          Phys.Rev. {\bf C33} (1986) 1740;
\item{18)}S. Ishikawa and T. Sasakawa, 
          Few-Body Syst. {\bf 1} (1986) 143;
          Y.Wu, S.Ishikawa and T.Sasakawa, 
          Few-Body Syst. {\bf 15} (1993) 145
\item{19)}Y.Wu, S.Ishikawa and T.Sasakawa, 
          Phys.Rev.Lett. {\bf 64} (1990) 1875;
          T.Sasakawa, S.Ishikawa, Y.Wu and T--Y.Saito,
          Phys.Rev.Lett. {\bf 68} (1992) 3503;
\item{20)}C.R.Chen, G.L.Payne, J.L.Friar and B.F.Gibson,
          Phys.Rev. {\bf C44}, (1991) 50
\item{21)}W.Gl\"ockle, H.Witala and T.Cornelius,
          Nucl.Phys. {\bf A508} (1990) 115c
\item{22)}D.Huber, H.Witala and W.Gl\"ockle, 
          Few Body Syst. {\bf 14} (1993) 171
\item{23)}E.O.Alt and M.Rauh {\sl in} : Proceedings of 14th European
          Conference on Few Body Problems in Physics, Amsterdam 1993
\item{24)}L.M.Delves in Advances in Nuclear Physics (Vol.5), p.126; 
          M.Baranger and E.Vogt (eds.), New York-London: Plenum Press 1972
\item{25)}C. R. Chen, G. L. Payne, J. L. Friar and B. F. Gibson, 
          Phys.Rev. {\bf C39} (1989) 1261;
\item{26)}L.M.Delves and M.A.Hennell,
          Nucl.Phys. {\bf A168} (1971) 347
\item{27)}R.A.Malfliet and J.A.Tjon, Nucl.Phys. {\bf A217}, (1969) 161
\item{28)} G.L.Payne, J.L.Friar and B.F.Gibson, Phys.Rev. {\bf C26}, 
          (1982) 1385;
          J.L.Friar, B.F.Gibson and G.L.Payne, Phys.Rev. {\bf C28}, 
          (1983) 983
\item{29)}R.G.Seyler, Nucl.Phys. {\bf A124} (1969) 253
\item{30)}A.Kievsky, M.Viviani and S.Rosati, 
          Nucl.Phys. {\bf A501} (1989) 503;
          A.Kievsky, M.Viviani and S.Rosati, 
          Few-Body Syst. {\bf 11} (1991) 111

\vfill\eject
\def\a{\alpha}

\def\r{\rho}

\def\O{\Omega} 

\def\he4{{}^4{\rm He}}

\def\qq{\quad}
\def\tablerule{\noalign{\hrule}}
\def\htm{{\hbar^2 \over m}}

\def\aa{{\alpha,\alpha'}}
\def\kk{{k,k'}}
\def\ap{{\alpha'}}
\def\Y{{\cal Y}}
%
\def\mev{\rm MeV}
\def\pd{$P_D$}
\def\pp{$P_P$}
\def\ps{$P_{S'}$}

\noindent{\bf Table Captions.}\bigskip
\noindent{Table 1.} Binding energy (B), kinetic energy mean value (T)
and $S'$--, $P$ and $D$--wave percentages for $^3$H and $^3$He in terms
of the number of channels $N_c$. The 26--channels calculations of ref.[3] are
reported in the last row for each case.
\bigskip
\noindent{Table 2.} Binding energies in terms of the number of channels
$N_c$ for the indicated potential models. The 34--channels CSF calculations
of ref.[17] are given in the last row for the sake of comparison.
\bigskip
\noindent{Table 3.} Results for the triton corresponding to $N_c=18$,
for the two potential model indicated. The $\pi$N form
factor is chosen to fit the experimental binding as is explained in the text.
\bigskip
\noindent{Table 4.} The doublet and quartet scattering lengths as a function
of the number of channels $N_c$. The 34--channel CSF calculations of 
ref.[20] are given in the last row for each case.
\bigskip
\noindent{Table 5.} First order and second order estimates for the reactance
matrix in three different approximations, as explained in the text.
The central column correspons to the evaluation of the
last term of eq.(3.10) which summed to the first order estimate gives
the second order estimate.
\bigskip
\noindent{Table 6.} Phase shifts and mixing parameters in degrees, 
calculated for
the indicated c.m. energies. For each energy the first and second column 
correspond to the n--d and p--d case, respectively.
The nuclear potential is
the AV14 and the numbers in parenthesis correspond to the AV14+BR model.
\vfill
\eject

\topinsert{
$$\vcenter{\offinterlineskip\hrule
\def\qq{\quad}
\def\Nc{$N_c$}
\def\spali{height5pt &\omit&&\omit&\omit&\omit&\omit&\omit&\cr}
\def\tablerule{\noalign{\hrule}}
\halign{& \vrule# 
        & \qq\hfil#\hfil\qq 
        & \vrule#  
        & \qq\hfil#\hfil\qq 
        & \qq\hfil#\hfil\qq 
        & \qq\hfil#\hfil\qq 
        & \qq\hfil#\hfil\qq 
        & \qq\hfil#\hfil\qq 
        & \vrule# \cr
\tablerule \spali 
&     &&\multispan5 \hfil $^3$H \hfil &\cr\spali
\tablerule \spali 
& \Nc && B(\mev )& T(\mev )& \ps (\%)& \pd (\%)& \pp (\%)&\cr\spali
\tablerule \spali 
& 8     && 7.660 & 45.551 & 1.128 & 8.926 & 0.066 &\cr\spali
& 12    && 7.678 & 45.645 & 1.127 & 8.962 & 0.076 &\cr\spali
& 18    && 7.683 & 45.671 & 1.126 & 8.965 & 0.076 &\cr\spali
&       &&       &        &       &       &       &\cr\spali
&ref.[2]&& 7.684 & 45.677 & 1.126 & 8.968 & 0.076 &\cr\spali
\tablerule \spali 
&     &&\multispan5 \hfil $^3$He\hfil &\cr\spali
\tablerule \spali 
& \Nc && B(\mev )& T(\mev )& \ps (\%)& \pd (\%)& \pp (\%)&\cr\spali
\tablerule \spali 
& 8     && 7.010 & 44.687 & 1.318 & 8.890 & 0.065 &\cr\spali
& 12    && 7.027 & 44.780 & 1.315 & 8.926 & 0.075 &\cr\spali
& 18    && 7.032 & 44.797 & 1.314 & 8.931 & 0.075 &\cr\spali
&       &&       &        &       &       &       &\cr\spali
&ref.[2]&& 7.033 & 44.812 & 1.314 & 8.932 & 0.075 &\cr\spali}
\hrule}$$\smallskip
\centerline{\bf Table~1.} 
}\endinsert
\bigskip\bigskip

\topinsert{
$$\vcenter{\offinterlineskip\hrule
\def\qq{\quad}
\def\Nc{$N_c$}
\def\spali{height5pt &\omit&&\omit&\omit&\cr}
\def\tablerule{\noalign{\hrule}}
\halign{& \vrule# 
        & \qq\hfil#\hfil\qq 
        & \vrule#  
        & \qq\hfil#\hfil\qq 
        & \qq\hfil#\hfil\qq 
        & \vrule# \cr
\tablerule \spali 
&     && AV14+TM & AV14+BR &\cr\spali
& \Nc && B(MeV)  & B(MeV)  &\cr\spali
\tablerule \spali 
& 8    && 9.241  & 9.153 & \cr\spali
& 10   && 9.306  & 9.218 & \cr\spali
& 12   && 9.315  & 9.225 & \cr\spali 
& 14   && 9.325  & 9.235 & \cr\spali 
&      &&        &       & \cr\spali
& CSF  && 9.32   & 9.22 & \cr\spali}
\hrule}$$\smallskip
\centerline{\bf Table~2.} 
}\endinsert
\bigskip\bigskip

\midinsert{
$$\vcenter{\offinterlineskip\hrule
\def\qq{\quad}
\def\spali{height5pt &\omit&&\omit&\omit&\omit&\omit&\omit&\cr}
\def\tablerule{\noalign{\hrule}}
\halign{& \vrule# 
        & \qq\hfil#\hfil\qq 
        & \vrule#  
        & \qq\hfil#\hfil\qq 
        & \qq\hfil#\hfil\qq 
        & \qq\hfil#\hfil\qq 
        & \qq\hfil#\hfil\qq 
        & \qq\hfil#\hfil\qq 
        & \vrule# \cr
\tablerule \spali 
& Potential && B(\mev )& T(\mev )& \ps (\%)& \pd (\%)& \pp (\%)&\cr\spali
\tablerule \spali 
& AV14+BR && 8.481 & 49.31  & 0.928 & 9.547 & 0.139 &\cr\spali
& AV14+TM && 8.480 & 49.30  & 0.938 & 9.255 & 0.161 &\cr\spali}
\hrule}$$\smallskip
\centerline{\bf Table~3.} 
}\endinsert
\bigskip\bigskip

\midinsert{
$$\vcenter{\offinterlineskip\hrule
\def\qq{\quad}
\def\spali{height5pt &\omit&&\omit&\omit&&\omit&\omit&\cr}
\def\tablerule{\noalign{\hrule}}
\def\s{\phantom{-}}
\halign{& \vrule# 
        & \qq\hfil#\hfil\qq 
        & \vrule#  
        & \qq\hfil#\hfil\qq 
        & \qq\hfil#\hfil\qq 
        & \vrule# 
        & \qq\hfil#\hfil\qq 
        & \qq\hfil#\hfil\qq 
        & \vrule# \cr
\tablerule \spali 
& $N_c$  &&\multispan2\hfil AV14     \hfil&&
           \multispan2\hfil AV14 + BR\hfil &\cr\spali
\tablerule \spali 
&        &&$^2a_{nd}$&$^2a_{pd}$ &&$^2a_{nd}$&$^2a_{pd}$&\cr\spali
\tablerule \spali 
&  3     &&          &           &&          &          &\cr\spali
&  8     && 1.211    & 0.980     &&\s0.048   & -1.056   &\cr\spali
&  10    && 1.198    & 0.957     && -0.003   & -1.139   &\cr\spali
&  12    && 1.196    & 0.954     && -0.010   & -1.145   &\cr\spali
\tablerule\spali
& CSF    && 1.204    & 0.965     && -0.001   & -1.136   &\cr\spali
\tablerule\spali
&        &&$^4a_{nd}$&$^4a_{pd}$ &&$^4a_{nd}$&$^4a_{pd}$&\cr\spali
\tablerule\spali
&  3     && 6.383    & 13.791    &&  6.376   &  13.830  &\cr\spali
&  10    && 6.382    & 13.785    &&  6.375   &  13.825  &\cr\spali
&  14    && 6.381    & 13.781    &&  6.374   &  13.820  &\cr\spali
&  18    && 6.380    & 13.779    &&  6.373   &  13.819  &\cr\spali
\tablerule\spali
& CSF    && 6.380    & 13.764    &&  6.381   &  13.765  &\cr\spali}
\hrule}$$\smallskip
\centerline{\bf Table~4.} 
}\endinsert
\bigskip\bigskip


\midinsert{
$$\vcenter{\offinterlineskip\hrule
\def\Rw{{}^J{\widetilde R}^{SS'}_{LL'}}
\def\a{{1\over 2}}
\def\b{{3\over 2}}
\def\Ra{{}^{1/2}{\widetilde R}^{\a\a}_{00}}
\def\Rb{{}^{1/2}{\widetilde R}^{\a\b}_{02}}
\def\Rc{{}^{1/2}{\widetilde R}^{\b\a}_{20}}
\def\Rd{{}^{1/2}{\widetilde R}^{\b\b}_{22}}
\def\st{1^{\underline{st}}}
\def\nd{2^{\underline{nd}}}
\def\cor{<\Psi_{L'S'J}\mid{\cal L}\mid\Psi_{LSJ}>}
\def\qq{\quad}
\def\spali{height5pt &\omit&&\omit&&\omit&\omit&\omit&\cr}
\def\tablerule{\noalign{\hrule}}
\halign{& \vrule# 
        & \qq\hfil#\hfil\qq 
        & \vrule#  
        & \qq\hfil#\hfil\qq 
        & \vrule#  
        & \qq\hfil#\hfil\qq 
        & \qq\hfil#\hfil\qq 
        & \qq\hfil#\hfil\qq 
        & \vrule# \cr
\tablerule \spali 
&$M(N_c)$&& $\Rw$ &&$\st-order$&  $\cor$  &$\nd-order$&\cr\spali
\tablerule \spali 
&  3(8)  && $\Ra$ && 2.778 & -0.003   & 2.775  &\cr\spali
&        && $\Rb$ && 0.821 &  0.028   & 0.849  &\cr\spali
&        && $\Rc$ && 0.850 & -0.001   & 0.849  &\cr\spali
&        && $\Rd$ && 62.07 &  3.665   & 65.74  &\cr\spali
\tablerule \spali 
&  6(8)  && $\Ra$ && 2.753 &  0.002   & 2.755  &\cr\spali
&        && $\Rb$ && 0.857 & -0.008   & 0.849  &\cr\spali
&        && $\Rc$ && 0.847 &  0.002   & 0.849  &\cr\spali
&        && $\Rd$ && 65.84 &  0.316   & 65.52  &\cr\spali
\tablerule \spali 
&  6(10) && $\Ra$ && 2.746 &  0.001   & 2.747  &\cr\spali
&        && $\Rb$ && 0.854 & -0.009   & 0.845  &\cr\spali
&        && $\Rc$ && 0.845 &  0.000   & 0.845  &\cr\spali
&        && $\Rd$ && 65.88 & -0.416   & 65.46  &\cr\spali}
\hrule}$$\smallskip
\centerline{\bf Table~5.} 
}\endinsert
\bigskip\bigskip


\midinsert{
$$\vcenter{\offinterlineskip\hrule
\def\eps{\varepsilon}
\def\zet{\zeta}
\def\qq{\quad}
\def\spali{height5pt &\omit&&\omit&\omit&&\omit&\omit&&\omit&\omit&\cr}
\def\tablerule{\noalign{\hrule}}
\halign{& \vrule# 
        & \qq\hfil#\hfil\qq 
        & \vrule#  
        & \qq\hfil#\hfil\qq 
        & \qq\hfil#\hfil\qq 
        & \vrule#  
        & \qq\hfil#\hfil\qq 
        & \qq\hfil#\hfil\qq 
        & \vrule#  
        & \qq\hfil#\hfil\qq 
        & \qq\hfil#\hfil\qq 
        & \vrule# \cr
\tablerule \spali 
& E$_{c.m.}$(MeV) &&\multispan2 \hfil 0.667\hfil 
                  &&\multispan2 \hfil 1.333\hfil 
                  &&\multispan2 \hfil 2.0  \hfil 
                                                    &\cr\spali
\tablerule \spali 
&            && n--d  & p--d  && n--d  & p--d  && n--d  & p--d  &\cr\spali
\tablerule \spali 
&$^2S_{1/2}$ && -17.7 & -12.6 && -27.9 & -23.6 && -34.9 & -31.4 &\cr\spali
&            &&(-14.5)&(-9.54)&&(-24.0)&(-19.8)&&(-30.6)&(-27.3)&\cr\spali
&$^4D_{1/2}$ && -1.00 & -0.79 && -2.58 & -2.28 && -3.91 & -3.62 &\cr\spali
&            &&(-1.00)&(-0.78)&&(-2.57)&(-2.28)&&(-3.90)&(-3.61)&\cr\spali
&$\eta_{1/2}$&&  1.04 &  1.19 &&  1.21 &  1.25 &&  1.26 &  1.26 &\cr\spali
&            && (1.50)& (1.85)&& (1.60)& (1.73)&& (1.61)& (1.67)&\cr\spali
\tablerule \spali 
&$^4S_{3/2}$ && -47.2 & -37.4 && -61.3 & -53.5 && -70.5 & -63.7 &\cr\spali
&$^2D_{3/2}$ &&  0.60 &  0.45 &&  1.55 &  1.36 &&  2.42 &  2.20 &\cr\spali
&$^4D_{3/2}$ && -1.08 & -0.84 && -2.77 & -2.46 && -4.22 & -3.91 &\cr\spali
&$\eps_{3/2}$&&  0.65 &  0.83 &&  0.72 &  0.79 &&  0.78 &  0.84 &\cr\spali
&$\zet_{3/2}$&& -0.11 & -0.09 && -0.23 & -0.20 && -0.37 & -0.32 &\cr\spali
&$\eta_{3/2}$&&  0.55 &  0.53 &&  1.01 &  0.98 &&  1.44 &  1.39 &\cr\spali 
\tablerule \spali 
&$^2D_{5/2}$ &&  0.57 &  0.45 &&  1.53 &  1.34 &&  2.38 &  2.17 &\cr\spali
&$^4D_{5/2}$ && -1.14 & -0.91 && -2.98 & -2.64 && -4.57 & -4.23 &\cr\spali
&$\eps_{5/2}$&& -0.29 & -0.37 && -0.30 & -0.34 && -0.31 & -0.35 &\cr\spali
\tablerule \spali 
&$^4D_{7/2}$ && -1.06 & -0.84 && -2.73 & -2.42 && -4.15 & -3.84 &\cr\spali
\tablerule \spali 
&$^2P_{1/2}$ && -4.54 & -3.61 && -7.66 & -6.83 && -9.51 & -8.85 &\cr\spali
&$^4P_{1/2}$ &&  12.2 &  9.30 &&  19.9 &  17.5 &&  23.9 &  22.0 &\cr\spali
&$\eps_{1/2}$&&  2.92 &  2.51 &&  3.95 &  3.48 &&  5.75 &  4.46 &\cr\spali
\tablerule \spali 
&$^2P_{3/2}$ && -4.51 & -3.58 && -7.57 & -6.76 && -9.34 & -8.72 &\cr\spali
&$^4P_{3/2}$ &&  14.2 &  10.9 &&  22.6 &  20.0 &&  26.2 &  24.5 &\cr\spali
&$\eps_{3/2}$&& -1.02 & -0.86 && -1.45 & -1.25 && -1.91 & -1.67 &\cr\spali
\tablerule \spali 
&$^4P_{5/2}$ &&  13.2 &  10.1 &&  21.5 &  18.9 &&  25.6 &  23.6 &\cr\spali}
\hrule}$$\smallskip
\centerline{\bf Table~6.} 
}\endinsert
\end